\pdfoutput=1

\documentclass[11pt,nofootinbib]{article}
\usepackage{amsmath,amssymb,mathtools}
\usepackage{graphicx}
\usepackage[colorlinks=true,linkcolor=blue,citecolor=blue,urlcolor=blue,linktocpage=true]{hyperref}
\usepackage{subfigure}
\usepackage{comment}
\usepackage{tensor}
\usepackage{cases}

\numberwithin{equation}{section}

\begin{document}

\providecommand{\abs}[1]{\lvert#1\rvert}
\providecommand{\bd}[1]{\boldsymbol{#1}}

\begin{titlepage}

\setcounter{page}{1} \baselineskip=15.5pt \thispagestyle{empty}

\begin{flushright}
SISSA 51/2016/FISI
\end{flushright}
\vfil


\bigskip
\begin{center}
 {\LARGE \textbf{Spontaneous Baryogenesis without Baryon Isocurvature}}
\vskip 15pt
\end{center}

\vspace{0.5cm}
\begin{center}
{\Large 
Andrea De Simone
and
Takeshi Kobayashi
}\end{center}

\vspace{0.3cm}

\begin{center}
\textit{SISSA and INFN Sezione di Trieste, Via Bonomea 265, 34136 Trieste, Italy}\\

\vskip 14pt
E-mail:
\texttt{\href{mailto:andrea.desimone@sissa.it}{andrea.desimone@sissa.it}}, 
\texttt{\href{mailto:takeshi.kobayashi@sissa.it}{takeshi.kobayashi@sissa.it}}

\end{center} 



\vspace{1cm}

\noindent
We propose a new class of spontaneous baryogenesis models that does not produce baryon isocurvature perturbations. The baryon chemical potential in these models is independent of the field value of the baryon-generating scalar, hence the scalar field fluctuations are blocked from propagating into the baryon isocurvature. We demonstrate this mechanism in simple examples where spontaneous baryogenesis is driven by a non-canonical scalar field. The suppression of the baryon isocurvature allows spontaneous baryogenesis to be compatible even with high-scale inflation.

\vfil

\end{titlepage}

\newpage
\tableofcontents

\section{Introduction}
\label{sec:intro}

The three basic ingredients for creating the asymmetry between baryons and
antibaryons from an initially symmetric state was laid out by
Sakharov~\cite{Sakharov:1967dj}. However, the third condition of a
deviation from thermal equilibrium can actually be traded for a breaking
of the $CPT$~symmetry. 
This is the idea of spontaneous baryogenesis~\cite{Cohen:1987vi}, which
typically invokes a scalar field derivatively coupled to the baryon
current in the form~$(\partial_\mu \phi) j_B^\mu $.
With such an interaction, the time derivative of a coherent
scalar~$\dot{\phi}$ spontaneously 
breaks~$CPT$ and shifts the spectrum of baryons relative to that of
antibaryons by an amount $\mu \propto \dot{\phi}$.
As a consequence, baryogenesis is allowed even in thermal equilibrium
if baryon number nonconserving processes occur rapidly.
This mechanism has been implemented in
various particle physics models,
e.g.,~\cite{Cohen:1988kt,Dine:1990fj,Cohen:1991iu,Dolgov:1994zq,Dolgov:1996qq,Chiba:2003vp,Carroll:2005dj,Kusenko:2014lra,Kusenko:2014uta,Daido:2015gqa}.

The scalar condensate~$\phi$ which drives spontaneous
baryogenesis can leave further imprints in the subsequent
cosmology~\cite{DeSimone:2016ofp}.
One such example is the baryon isocurvature
perturbation~\cite{Turner:1988sq}, since
super-horizon field fluctuations of~$\phi$ sourced during inflation
give rise to spatial fluctuations of the baryon-to-photon ratio.
Hence the current observational bounds on isocurvature perturbations
provide strong constraints on spontaneous baryogenesis scenarios.
Here, the amplitude of the field fluctuation is set by
the scale of inflation,
therefore the bound on the baryon isocurvature can be translated into an upper
bound on the inflation scale.
In particular, for the minimal spontaneous baryogenesis scenario where
the scalar~$\phi$ possesses a quadratic potential,
the bound on baryon isocurvature from measurements of the cosmic
microwave background (CMB)  
constrains the inflationary Hubble rate as
$H_{\mathrm{inf}} \lesssim 10^{12}\, \mathrm{GeV}$~\cite{DeSimone:2016ofp}.
This implies that if inflationary gravitational waves are detected in the near
future, thus confirming an inflation scale higher than $10^{12}\,
\mathrm{GeV}$,
then the minimal scenario of spontaneous baryogenesis would be ruled out. 
As there are a variety of ongoing and upcoming experiments in search of
primordial gravitational waves,
it is of great interest to investigate whether high-scale inflation
rules out spontaneous baryogenesis in general.

The production of the baryon isocurvature in spontaneous baryogenesis is due
to the fact that the
relative shift between the baryons and antibaryons is set by the scalar
velocity, $\mu \propto \dot{\phi}$,
which takes slightly different values among different patches of the
universe in the presence of the scalar field fluctuations.
However, there are situations where the fluctuations in the field value
do not necessarily lead to fluctuations in the field velocity.
It was pointed out in~\cite{DeSimone:2016ofp} that for cases
where the scalar possesses non-quadratic potentials with inflection
points, the fluctuations in the scalar velocity is suppressed if the 
initial position of the scalar field happens to lie close
to an inflection point.
We should also mention that there have been other proposals which may
evade the isocurvature constraint; these include
compensating the baryon isocurvature with cold dark matter
isocurvature if the scalar serves as dark 
matter~\cite{DeSimone:2016ofp},
stabilizing the scalar in a false vacuum during inflation~\cite{Kusenko:2014lra},
driving spontaneous baryogenesis by domain walls~\cite{Daido:2015gqa},
or by the derivative of the Ricci scalar instead of a scalar
field~\cite{Davoudiasl:2004gf}.

In this paper we propose a new class of spontaneous baryogenesis models
that is free from baryon isocurvature perturbations.
The basic idea is to block the field fluctuations of the
baryon-generating scalar~$\phi$ from propagating into the baryon isocurvature
by invoking a combination of a derivative coupling~$f(\phi)
(\partial_\mu \phi) j_B^\mu$ and a scalar potential~$V(\phi)$
that renders the product~$f(\phi) V'(\phi)$ constant. 
Then, since the scalar slowly rolling along the potential possesses a
velocity of $\dot{\phi} \propto -V'(\phi)$, the induced 
shift in the baryon/antibaryon spectra,
\begin{equation}
 \mu \propto f(\phi) \dot{\phi} \propto f(\phi) V'(\phi) = \mathrm{const.},
\label{general}
\end{equation}
is independent of the scalar field value.
Therefore the resulting baryon number becomes spatially homogeneous
even though the scalar itself possesses field fluctuations.

We obtain a set of reasonably simple models where the situation 
of~(\ref{general}) is realized, by considering a scalar field with a
non-canonical kinetic term.
We discuss various special properties of such models,
and in particular we will find that a successful
baryogenesis is allowed even with high-scale inflation. 
Here, let us remark that the inflection point model
of~\cite{DeSimone:2016ofp}
also satisfies the condition~(\ref{general}) at special points
along the scalar potential.
However in the new class of models presented in this paper,
the suppression happens generically without the need of tuning the
initial position of the scalar. 
Furthermore, the striking feature that the baryon asymmetry is
independent of the scalar field value provides predictive power to the model.

The paper is organized as follows.
We first briefly explain how the condition~(\ref{general}) can be satisfied by
a non-canonical scalar in Section~\ref{sec:non-canonical}.
Then we study in detail two example models in Sections~\ref{sec:frac}
and~$\ref{sec:lin}$.
We conclude in Section~\ref{sec:conc}.

\section{Models with Non-Canonical Scalars}
\label{sec:non-canonical}

We begin by briefly describing a class of models where a non-canonical
scalar field drives spontaneous baryogenesis 
without producing baryon isocurvature. 
The Lagrangian of the model consists of a scalar with a non-canonical
kinetic term, a mass term, and a coupling to the divergence of a baryon
current of the form 
\begin{equation}
 S = \int d^4 x \sqrt{-g}
  \left[
   -\frac{1}{2} \sqrt{ 1 + \left(\frac{\phi }{\lambda}\right)^{2 n} } \, 
   g^{\mu  \nu}  \partial_\mu \phi \partial_\nu \phi
   - \frac{1}{2} m^2 \phi^2
   - \left( \frac{\phi }{f} \right)^n \nabla_\mu j_{B}^\mu
	  \right],
\label{Ln}
\end{equation}
where $n$ is a positive integer, while $\lambda$, $f$, and $m$ are mass scales. 
For small field values $\abs{\phi} \ll \lambda $, the kinetic term is
almost canonical.
On the other hand in the large field limit of $\phi \gg \lambda $, the kinetic term 
becomes approximately proportional to $\phi^n (\partial \phi)^2$ and
thus it can be made canonical by redefining the field as
\begin{equation}
 \sigma \propto \phi^{\frac{n+2}{2}}.
\end{equation}
In terms of this canonically normalized field,
the originally mass term serves as a power-law potential of
\begin{equation}
 V = \frac{1}{2} m^2 \phi^2 \propto \sigma^{\frac{4}{n+2}}.
\end{equation}
Hence the time derivative of the scalar that slowly rolls along
this potential depends on the field value as
\begin{equation}
 \dot{\sigma} \propto - \frac{\partial V}{\partial \sigma } \propto
  \sigma^{- \frac{n-2}{n+2}} .
  \label{1.4}
\end{equation}
The coupling to the baryon current, after integration by parts, can also
be rewritten in terms of the canonical field as
\begin{equation}
 - \sqrt{-g} \left( \frac{\phi }{f} \right)^n \nabla_\mu j_{B}^\mu
  \Rightarrow
  \sqrt{-g} \left\{\partial_\mu \left( \frac{\phi }{f} \right)^n \right\}
j_B^\mu 
  \propto 
  \sigma^{\frac{n-2}{n+2}} (\partial_\mu \sigma) j_B^\mu.
\end{equation}
Here one sees that the condition~(\ref{general}) is satisfied for the
canonical field, therefore
the shift in the baryon/antibaryon spectra is independent of the
field value,
\begin{equation}
 \mu \propto \sigma^{\frac{n-2}{n+2}} \dot{\sigma} \propto \sigma^0.
  \label{mu-scaling}
\end{equation}
Thus it is clear that the field fluctuations of~$\sigma$ do not source
fluctuations in the baryon number. 

This mechanism of suppressing the baryon isocurvature is simplest to
understand in the case of $n = 2$:
Here the potential in the large field limit is linear, $V \propto \sigma$,
whose constant tilt renders the scalar velocity~$\dot{\sigma}$ homogeneous.
Since the derivative coupling also takes a linear form~$(\partial_\mu \sigma)
j_B^{\mu}$, the spectrum shift is simply $\mu \propto \dot{\sigma}$,
which is guaranteed to be homogeneous.
The suppression of the baryon isocurvature in linear potentials was also
pointed out in~\cite{DeSimone:2016ofp},
however, as is clear from the above discussion, the suppression happens
for an arbitrary positive integer~$n$.

In the following sections, 
we study in detail the model~(\ref{Ln}) and its variant,
focusing on cases of $n = 1$ and $2$. 
In addition to the suppression of the baryon isocurvature, 
the scalar undergoes nonstandard evolution after baryogenesis,
and as a consequence the model behaves quite differently from the usual
spontaneous baryogenesis scenarios.

\section{$n = 1$ : Spontaneous Baryogenesis with Fractional Power Terms}
\label{sec:frac}

In this section we analyze spontaneous baryogenesis with the case of $ n = 1$
in~(\ref{Ln}), where the Lagrangian of a real scalar~$\phi$ reads
\begin{equation}
 S = \int d^4 x \sqrt{-g}
  \left[
   -\frac{1}{2} \sqrt{ 1 + \left(\frac{\phi }{\lambda }\right)^{2} } \, 
   g^{\mu  \nu}  \partial_\mu \phi \partial_\nu \phi
   - \frac{1}{2} m^2 \phi^2
   - \frac{\phi }{f} \sum_i c_i
   \nabla_\mu j_{i}^\mu
  \right].
\label{Sn1}
\end{equation}
Here $\lambda $, $m$, $f$ are mass scales, $c_i$ is a dimensionless
coefficient, and $j_i^\mu$ represents the current of a
particle/antiparticle pair~$i$ which has baryon number~$(-)B_i$
for the (anti)particle.
The time component $j_i^0 = n_i - \bar{n}_i$ represents 
the difference in the number density between the particle and
antiparticle, and the sum~$\sum_i$ runs over all particle species
coupled to~$\phi$.

Let us briefly comment on the possible origin of the above action.
A nice example where the non-canonical kinetic term can arise is
through the Nambu--Goto action of a brane, as in the D-brane monodromy
inflation models of~\cite{Silverstein:2008sg,McAllister:2008hb}.
See also \cite{Takahashi:2010ky} which invoked similar
kinetic terms in the context of inflation.
Regarding the coupling to the divergence of the current,
such a term can arise, for instance, from anomalous couplings to the SU(2) gauge
fields. (However in such cases the coupling term would only be
effective when sphalerons are in
equilibrium~\cite{Daido:2015gqa,Shi:2015zwa}.) 
The mass scale~$f$ in the coupling can be related to the scale of new
physics;
for example, in the models
of~\cite{Cohen:1988kt,Dolgov:1994zq,Dolgov:1996qq}, the scalar~$\phi$ 
is a pseudo-Nambu--Goldstone boson of the broken baryon number,
with $f$ being the associated symmetry breaking scale.
Although in this paper we do not specify the origin of the scalar and
its coupling to the baryon current,
we will restrict our attention to field ranges of~$\phi$ that do
not exceed~$f$.
Furthermore, we assume~$f$ to be larger than the Hubble expansion rate and 
the cosmic temperature during the relevant epochs.
Some more discussions on the validity of the effective
field theory will be provided in Section~\ref{subsec:cutoff}.

As for the scale~$\lambda$ in the kinetic term, we suppose it to follow
a hierarchy of $\lambda \ll f$. 
In the small field limit of~$\abs{\phi} \ll \lambda $, the field~$\phi$ is
almost canonical and the system reduces to the usual case studied in
most spontaneous baryogenesis models.
On the other hand, in order to study the large field regime of
$\lambda  \ll \abs{\phi} \lesssim f$, let us 
focus on positive field values $\phi > 0$ and introduce a new field,
\begin{equation}
 \sigma = \frac{2}{3} \frac{\phi^{3/2}}{\lambda^{1/2}},
\end{equation}
with which the action is rewritten as
\begin{multline}
 S = \int d^4 x \sqrt{-g}
  \Biggl[
-\frac{1}{2} \sqrt{1 + \left( \frac{2}{3} \frac{\lambda }{\sigma }
		       \right)^{4/3}}  g^{\mu \nu} \partial_\mu \sigma
\partial_\nu \sigma
- \frac{1}{2} \left( \frac{3}{2}\right)^{4/3} m^2 \lambda^{2/3} \sigma^{4/3}
 \\
 + \left( \frac{2}{3} \frac{\lambda }{\sigma }\right)^{1/3} 
\frac{\partial_\mu \sigma }{f }
\sum_i c_i j_i^{\mu}
  \Biggr].
\label{eq3.3}
\end{multline}
Here, for the coupling term to the current, we have performed an
integration by parts and dropped the total derivative. 
In the large field regime of $\phi \gg \lambda$, 
or $\sigma \gg \lambda$, the almost canonical field is~$\sigma$
and it has a potential with a fractional power,
\begin{equation}
 V(\sigma)  = \frac{1}{2} \left( \frac{3}{2}\right)^{4/3} m^2 \lambda^{2/3}
  \sigma^{4/3},
  \label{V4over3}
\end{equation}
as well as a fractional power-law coupling to the current.

\subsection{Qualitative Picture}

\begin{figure}[t]
  \begin{center}
  \begin{center}
  \includegraphics[width=0.7\linewidth]{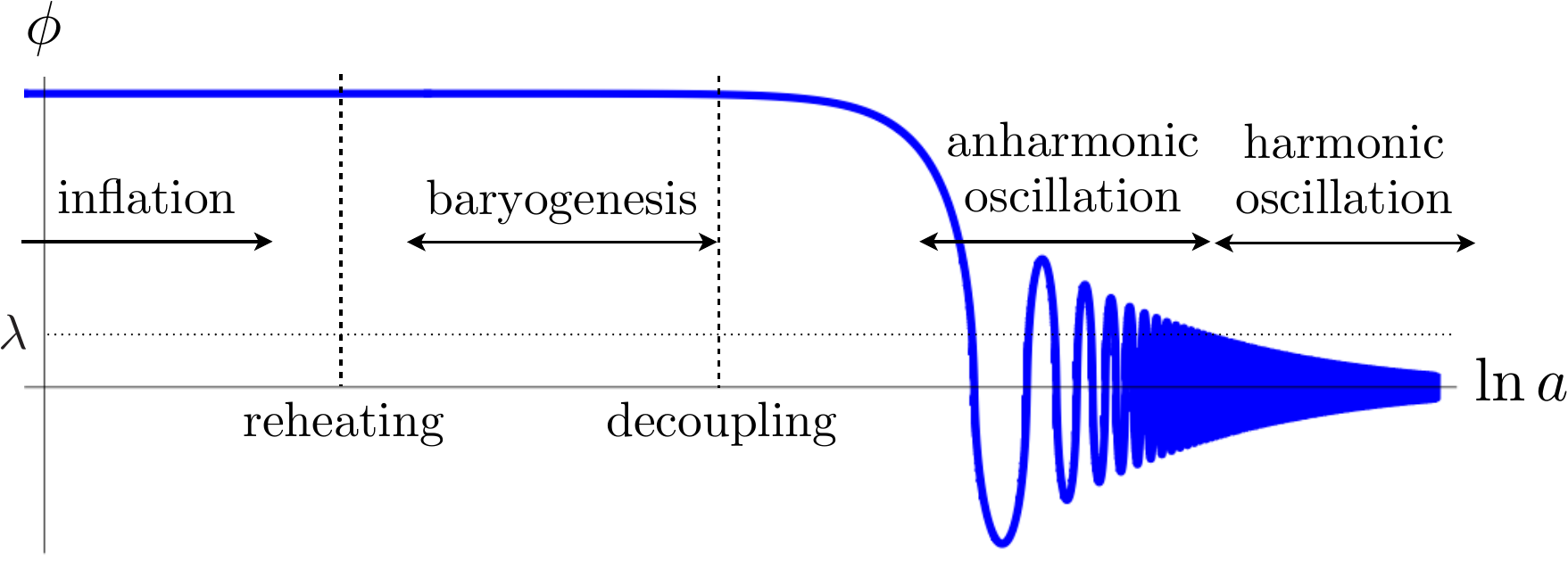}
  \end{center}
  \caption{Schematic of the scalar field dynamics (not to scale).}
  \label{fig:schematic}
  \end{center}
\end{figure}

Now let us analyze spontaneous baryogenesis with the above model.
We illustrate the cosmological history and the scalar field dynamics in
Figure~\ref{fig:schematic}:
Some time after cosmic inflation, the universe undergoes reheating.
Then supposing some baryon number nonconserving processes to be in
equilibrium, the baryon asymmetry is produced as the scalar field rolls
along its potential.
The baryon number eventually freezes in when the baryon number
nonconserving processes fall out of equilibrium.
After baryogenesis, as the Hubble friction becomes weaker, the scalar
starts to oscillate about the minimum of its potential.

We suppose the scalar field to be initially located in the large field
regime: 
\begin{equation}
 \lambda \ll \phi_{\star} \lesssim f,
  \label{large-field}
\end{equation}
where we have used~$\phi_\star$ to denote the scalar field value during
inflation when the CMB pivot scale~$k_\star$ leaves the horizon. 
Then, as we will explicitly show, 
the oscillation amplitude~$\bar{\phi}$ at the beginning of the
oscillation is larger than~$\lambda$ and hence the scalar initially undergoes
anharmonic oscillations along the fractional power-law potential~(\ref{V4over3}).
After $\bar{\phi}$ becomes smaller than~$\lambda$, the oscillation becomes
harmonic.

\subsection{Scalar Field Dynamics}

The universe is considered to initially undergo
inflation, then to be effectively matter-dominated (MD), and after
reheating to be radiation-dominated (RD).
Spontaneous baryogenesis is supposed to happen during the RD epoch.
We describe the cosmological background in terms of a flat FRW universe,
\begin{equation}
 ds^2 = -dt^2 + a(t)^2 d \bd{x}^2,
\end{equation}
whose Hubble expansion rate follows
\begin{equation}
 \frac{\dot{H}}{H^2} = -\frac{3 (1 + w)}{2},
\label{Hdot}
\end{equation}
with $w = -1$, $0$, and $1/3$ during the inflation, MD, and RD epochs,
respectively.
An overdot denotes a derivative in terms of the cosmological time~$t$.

The scalar field is considered to have a negligible effect on the expansion
of the very early universe.
Therefore we require the initial field value of the
canonical field to be sub-Planckian, 
\begin{equation}
 \sigma_\star < M_p,
  \quad
  \mathrm{i.e.,}
  \quad
\phi_\star < \left(\frac{3}{2}\right)^{2/3} M_p^{2/3}\lambda^{1/3},
\label{sub-Planck}
\end{equation}
otherwise the scalar could dominate the universe. 
The equation of motion of a homogeneous scalar field, i.e. $\sigma =
\sigma(t)$, reads
\begin{equation}
 \gamma(\sigma)
  (\ddot{\sigma} + 3 H \dot{\sigma})
  - \frac{1}{2 \gamma (\sigma)} \left(\frac{2}{3}\right)^{7/3}
\frac{\lambda^{4/3}
  \dot{\sigma}^2}{\sigma^{7/3}}
  + V'(\sigma) +
  \left(\frac{2}{3} \frac{\lambda }{\sigma } \right)^{1/3}
  \frac{\sum_i c_i \nabla_\mu j_i^{\mu} }{f}  = 0,
\label{sigmaEoM}
\end{equation}
where $\gamma (\sigma)$ is defined as
\begin{equation}
 \gamma (\sigma) =
  \sqrt{ 1 + \left( \frac{2}{3} \frac{\lambda  }{\sigma }  \right)^{4/3}} .
\end{equation}
The scalar field dynamics prior to the onset of the oscillations is
approximately given by
\begin{equation}
 \frac{3 (3+w)}{2} H \dot{\sigma} \simeq -V'(\sigma).
\label{slow-vary}
\end{equation}
One can check that this expression provides a
good approximation to the equation of motion~(\ref{sigmaEoM}) when
\begin{equation}
   \left( \frac{\lambda }{\sigma } \right)^{4/3}, \, \, 
    \frac{V''(\sigma)}{H^2},  \, \, 
  \frac{\abs{\sum_i c_i \nabla_\mu j_i^\mu }}{m^2 \lambda^{1/3} f \sigma
  ^{2/3}}
  \ll 1.
  \label{SVcond}
\end{equation}
These conditions can be understood as the requirements of, respectively,
large field value $\phi^2 \gg \lambda^2$,
small effective mass compared to the Hubble rate,
and negligible backreaction from the baryons during baryogenesis.
We also note that the approximation~(\ref{slow-vary}) is actually an attractor
while the conditions~(\ref{SVcond}) are satisfied;
see e.g. the analyses in Appendix~A of~\cite{Kawasaki:2011pd}.

Here we have discussed a homogeneous scalar,
however we remark that since the scalar's effective mass is
initially much lighter than the Hubble rate (cf.~(\ref{SVcond})),
the scalar field actually obtains spatial fluctuations
on super-horizon scales during inflation.
When discussing cosmological perturbations in the following sections,
we will take into account the scalar field fluctuations by
noting that the initial field value during inflation is slightly
different among different patches of the universe
by $\delta \sigma \sim H_{\mathrm{inf}} / 2 \pi$.
We also note that throughout this paper we will focus on cases where the field fluctuations
can be treated as small perturbations, hence we suppose
\begin{equation}
  \sigma_\star >  H_{\mathrm{inf}},
  \quad
  \mathrm{i.e.,}
  \quad
\phi_\star > \left(\frac{3}{2}\right)^{2/3} H_{\mathrm{inf}}^{2/3}\lambda^{1/3},
\label{super-Hinf}
\end{equation}
for the scalar field value during inflation.

\subsection{Baryon Asymmetry}

The coherent background of~$\dot{\phi}$ spontaneously breaks the $CPT$
symmetry, and thus sources a relative shift in the energy spectra of the
particles and antiparticles through the coupling term~$(\partial_\mu
\phi / f) c_i j_i^\mu$.
When the (anti)particles~$i$ are in thermal equilibrium, the energy shift
can be interpreted as particles obtaining an effective chemical potential of
\begin{equation}
 \mu_i = -c_i \frac{\dot{\phi}}{f} =
  - c_i \left( \frac{2}{3} \frac{\lambda }{ \sigma } \right)^{1/3}
  \frac{\dot{\sigma} }{f}
  = \frac{c_i}{5} \frac{m^2 \lambda }{f H},
  \label{chpot}
\end{equation}
and $-\mu_i$ for antiparticles.
In the far right hand side, we used the slow-varying
approximation for the scalar velocity~(\ref{slow-vary}) 
in a RD universe ($w = 1/3$).
Here one clearly sees that the chemical potential is independent of
the scalar field value~$\sigma$, as was discussed
around~(\ref{mu-scaling}). 
The chemical potential gives rise to a baryon asymmetry,
when there are baryon violating processes occurring rapidly. 
Supposing all the particle species~$i$ to be relativistic
fermions and ignoring their masses,
the difference in the number densities
of the particles and antiparticles is 
\begin{equation}
 j_i^0 = n_i - \bar{n}_i =
  \frac{g_i}{6} \mu_i T^2
  \left\{ 1 + \mathcal{O} \left( \frac{\mu_i}{T} \right)^2 \right\},
\label{ji0}
\end{equation}
where $g_i$ represents the internal degrees of freedom of the
(anti)particle~$i$, and we have assumed the chemical potential to be much
smaller than the cosmic temperature, i.e. $\mu_i^2  \ll  T^2 $. 
Hence the ratio between the baryon number density
$ n_B = \sum_i B_i (n_i - \bar{n}_i)$
and the entropy density
\begin{equation}
 s =  \frac{2 \pi^2}{45} g_{s*} T^3
  \label{entden}
\end{equation}
is obtained as
\begin{equation}
 \frac{n_B}{s} = \frac{15 }{4 \pi^2 }
  \frac{\sum_i B_i g_i \mu_i}{g_{s*} T}.
  \label{b2p-rat}
\end{equation}
This ratio freezes in
after the baryon violating interactions fall out of equilibrium,
given that there are no further baryon or entropy production afterwards.
Hence the ratio at the decoupling of the baryon violating
interactions~$(n_B/s)_{\mathrm{dec}}$ should coincide with the present
value $(n_B / s)_0 \approx 8.6 \times 10^{-11}$ measured by {\it
Planck}~\cite{Ade:2015xua}.
Hereafter we denote the decoupling temperature by~$T_{\mathrm{dec}}$,
and also use the subscript~``dec'' for quantities measured at
decoupling.
On the other hand for quantities in the present universe, we use the 
subscript~``$0$''.
(Here one also sees from (\ref{b2p-rat}) that
the assumption of $\mu_{i, \mathrm{dec}}^2 \ll T_{\mathrm{dec}}^2$
is justified in the case of $(n_B/s)_{\mathrm{dec}} \approx 8.6 \times 10^{-11}$,
unless $\sum_i B_i g_i/g_{s*}$ takes an extremely small value.)

Using (\ref{chpot}) and 
the relation between the Hubble rate and the temperature in a RD universe:
\begin{equation}
 3 M_p^2 H^2 = \rho_{\mathrm{r}} =
  \frac{\pi^2}{30} g_* T^4, 
\label{RDuniv}
\end{equation}
the baryon-to-entropy ratio (\ref{b2p-rat}) at decoupling is obtained as
\begin{equation}
 \left. \frac{n_B}{s} \right|_{\mathrm{dec}} =
\frac{9 }{ \pi^3 } \left( \frac{5}{8}  \right)^{1/2}
  \frac{\sum_i B_i c_i g_i }{
g_{*\mathrm{dec}}^{1/2} \,  g_{s*\mathrm{dec}} }
\frac{M_p m^2 \lambda }{T_{\mathrm{dec}}^3 \,  f}.
  \label{2.21}
\end{equation}
This result should be contrasted to the baryon asymmetry created in
usual spontaneous baryogenesis scenarios
where $n_B/s$ is proportional to some powers of the scalar field
value~$\sigma$, and thus carries isocurvature perturbations of 
$\delta n_B / n_B \sim \delta \sigma / \sigma  \sim H_{\mathrm{inf}} / (2 \pi \sigma)$.
In our case, as long as the scalar follows the slow-varying
solution~(\ref{slow-vary}) during baryogenesis,
the $\sigma$~dependence drops out of the baryon number.
As a consequence, the inhomogeneities in~$\sigma$ are blocked from propagating
into those of the baryons,
and thus the baryon isocurvature perturbation is strongly suppressed.

However, we should also remark that the error in the slow-varying
approximation~(\ref{slow-vary}) can source subleading contributions to
the baryon asymmetry, which can produce a tiny but non-vanishing baryon
isocurvature.
We will estimate this effect numerically later when we
discuss the parameter space of the model.
Let us further note that, if the scalar ever dominates the
universe after creating the baryon asymmetry, then this would also lead
to baryon isocurvature perturbations. This issue will be discussed in
Section~\ref{subsec:curvaton-like}. 

Before closing this subsection, let us rewrite the requirement of
negligible backreaction from the 
baryons, i.e. the third of the slow-varying conditions~(\ref{SVcond}),
using (\ref{chpot}), (\ref{ji0}), and (\ref{RDuniv}).
Ignoring the spatial components of~$j_i^\mu$ and the time derivative
of~$g_*$, the condition is rewritten as
\begin{equation}
 \frac{\sum_{i} c_i^2 g_i}{10}
  \left( \frac{\lambda }{\sigma } \right)^{2/3}
  \left( \frac{T}{f} \right)^2
  \ll 1.
\label{HAinT}
\end{equation}
Hence we see that, as we have been assuming $f > T$,
the backreaction from the baryons is
guaranteed to be negligible in the large field regime $\sigma \gg
\lambda$ (unless $\sum_{i} c_i^2 g_i$ takes a large value).

\subsection{The Fate of the Scalar}

\subsubsection*{Onset of Oscillations}

Up until the end of baryogenesis, the scalar velocity~$\dot{\sigma}$ is
nonzero but the field value itself is effectively frozen.
This is clearly seen from the slow-varying approximation~(\ref{slow-vary}) giving
\begin{equation}
 \left| \frac{\dot{\sigma}}{H \sigma } \right|
  = \frac{2}{3+w} \frac{V''(\sigma) }{H^2},
\end{equation}
which is much smaller than unity while $V'' \ll H^2$.
However after the decoupling, as the Hubble friction becomes weaker, the
scalar eventually starts to oscillate about its potential minimum.
Note here that if the oscillation starts prior to decoupling,
the baryon asymmetry would be extremely suppressed 
as the effective chemical 
potential becomes tiny when averaged over the oscillations.
(See \cite{Dolgov:1994zq,Dolgov:1996qq} and 
Appendix of~\cite{DeSimone:2016ofp} for detailed discussions on this.) 
Thus we require the scalar to start the oscillation after decoupling,
and discuss the fate of the oscillating scalar condensate in this subsection.

We begin by defining the `onset' of the scalar oscillation as when the field
excursion during one Hubble time becomes comparable to the distance to
the potential minimum, i.e.,
\begin{equation}
 \left| \frac{\dot{\sigma }}{H \sigma }  \right|_{\mathrm{osc}} = 1,
\end{equation}
and denote quantities measured at this moment by the subscript~``osc''.
Then by supposing the slow-varying approximation~(\ref{slow-vary}) to be
valid until the onset of the oscillation, we obtain
\begin{equation}
 H_{\mathrm{osc}}^2 = \frac{1}{5} \left(\frac{3}{2} \right)^{1/3}
  \left( \frac{\lambda}{\sigma_{\mathrm{osc}}} \right)^{2/3} m^2,
\label{Hosc}
\end{equation}
where we used $w = 1/3$
as we are interested in cases where the scalar starts its
oscillation during the RD era. 

The field value at the onset of the oscillation~$\sigma_{\mathrm{osc}}$
can further be expressed in terms of field values at earlier times.
For this purpose, let us 
rewrite the slow-varying solution~(\ref{slow-vary}) as 
$d \sigma / V'(\sigma) = -2 d t / \{ 3 (3+w) H\}$,
and integrate both sides through the inflation, MD, and RD epochs
using (\ref{Hdot});
integrating from when the pivot
scale~$k_{\star}$ exits the horizon during inflation
until the onset of the oscillation, one obtains
\begin{equation}
 \left(\frac{3}{2} \frac{\sigma_\star}{\lambda } \right)^{2/3}
  -  \left(\frac{3}{2} \frac{\sigma_{\mathrm{osc}}}{\lambda } \right)^{2/3}
  = \frac{\mathcal{N}_\star}{3} \frac{m^2}{H_{\mathrm{inf}}^2}
+ \frac{2}{27} \left( \frac{m^2}{H_{\mathrm{reh}}^2 }- 
\frac{m^2}{  H_{\mathrm{inf}}^2 } \right)
+ \frac{1}{20} \left( \frac{m^2}{H_{\mathrm{osc}}^2 }- 
\frac{m^2}{  H_{\mathrm{reh}}^2 } \right).
\label{2.24}
\end{equation}
Here, $\mathcal{N}_\star$ denotes the number of $e$-folds between the
horizon exit of~$k_{\star}$ and the end of inflation,
and $H_{\mathrm{reh}}$ denotes the Hubble rate at reheating. 
Hence, supposing a hierarchy between the Hubble rates as
\begin{equation}
 H_{\mathrm{osc}}^2 \ll H_{\mathrm{reh}}^2, \, \, 
  \frac{3}{20 \mathcal{N}_\star} H_{\mathrm{inf}}^2 ,
\label{2.26}
\end{equation}
and using the expression~(\ref{Hosc}) for~$H_{\mathrm{osc}}$,
one finds
\begin{equation}
 \sigma_{\mathrm{osc}} =
  \left(\frac{6}{7} \right)^{3/2} \sigma_\star,
  \quad
  \mathrm{i.e.,}
  \quad
  \phi_{\mathrm{osc}} = \frac{6}{7} \phi_\star. 
\label{phi_osc}
\end{equation}
This explicitly shows that, given that the scalar is located in the
large field regime during inflation, i.e. $\phi_\star \gg \lambda$,
then the oscillation also starts with an amplitude $\phi_{\mathrm{osc}}
\gg \lambda$ so that the oscillation is initially anharmonic.\footnote{The
field excursion during the $\mathcal{N}_\star$~$e$-foldings in the
inflation epoch, $\Delta \phi_\star$, can be computed by integrating
$d \sigma / V'(\sigma) = -d t / (3 H_{\mathrm{inf}})$ as
\begin{equation}
 \frac{\Delta \phi_\star}{\phi_\star} =
 \frac{\mathcal{N}_\star}{3} \frac{m^2 \lambda }{H_{\mathrm{inf}}^2
 \phi_\star}  
=  \frac{20   \mathcal{N}_\star}{21}
  \left( \frac{H_{\mathrm{osc}}}{H_{\mathrm{inf}}} \right)^2,
\end{equation}
where we also used (\ref{Hosc}) and (\ref{phi_osc}) upon obtaining the
far right hand side. 
Hence one sees that the condition of $H_{\mathrm{osc}}^2 \ll 3
H_{\mathrm{inf}}^2 / (20 \mathcal{N}_\star)$ in~(\ref{2.26}) 
implies that the scalar field is effectively frozen
during inflation.
If this condition is violated, then the scalar may start oscillating
before the end of inflation.
We also note that this condition is similar to the second of the
slow-varying condition (\ref{SVcond}) during inflation, but with an
additional factor of~$\mathcal{N_\star}$.}

One can also rewrite the requirement
that the scalar should start its oscillation after decoupling 
by using (\ref{Hosc}) and (\ref{phi_osc}) as
\begin{equation}
 \left(
\frac{H_{\mathrm{osc}}}{H_{\mathrm{dec}}}
 \right)^2
 = \frac{63 }{2 \pi^2 } \frac{1}{g_{*\mathrm{dec}}}
 \frac{M_p^2 m^2 \lambda }{T_{\mathrm{dec}}^4 \phi_{\star} } < 1,
\label{cond-ii}
\end{equation}
where we expressed the Hubble rate at decoupling in terms of the
temperature.

\subsubsection*{Scalar Abundance}

Let us now compute the energy density of the oscillating scalar field.
During the anharmonic oscillations along
the fractional power-law potential\footnote{We have defined
$\sigma$ to describe the $\phi > 0$ regime,
but one can similarly introduce an almost canonical field for the large
field regime in the negative side $\phi < 0$.}
$V\propto \sigma^{4/3}$, the scalar density 
redshifts as\footnote{A canonical scalar field that coherently
oscillates (mostly) along a 
power-law potential $V \propto \varphi^s$
can be described as a perfect fluid with an equation of state
parameter $ w = (s-2)/(s+2)$ when averaged over the oscillation.
Hence its energy density redshifts as
$\rho_{\varphi} \propto a^{-6s/(s+2)}$ in an expanding universe.}
\begin{equation}
 \rho_\phi \propto a^{-12/5},
\end{equation}
and thus the oscillation amplitude damps as
\begin{equation}
 \bar{\sigma} \propto a^{-9/5},
    \quad
  \mathrm{i.e.,}
  \quad
  \bar{\phi} \propto a^{-6/5}.
  \label{anh_amp_damp}
\end{equation}
When the amplitude becomes as small as $\bar{\phi} \lesssim \lambda$,
the scalar undergoes harmonic oscillations along the quadratic
potential, hence the scalings become
\begin{equation}
 \rho_\phi \propto a^{-3},
  \quad
  \bar{\phi} \propto a^{-3/2}.
  \label{har_amp_damp}
\end{equation}

Let us denote the time when the amplitude becomes $\bar{\phi} = \lambda$
by~$t_{\mathrm{q}}$, and quantities measured at
this time by the subscript~``q''.
We assume the time~$t_{\mathrm{q}}$ to be during the RD epoch.
While the scalar undergoes anharmonic oscillations,
the relation between the entropy density and the scalar amplitude is 
obtained from~(\ref{anh_amp_damp}) as 
$s \propto a^{-3} \propto \bar{\phi}^{5/2}$, yielding
\begin{equation}
  \frac{s_{\mathrm{q}}}{s_{\mathrm{osc}}} 
  = \left( \frac{\lambda }{\phi_{\mathrm{osc}}}\right)^{5/2}.
\label{sqsosc}
\end{equation}
This, combined with the harmonic redshifting~(\ref{har_amp_damp}) and
$\rho_{\phi\, \mathrm{q}} \simeq m^2 \lambda^2 / 2$ gives
the scalar density during the harmonic oscillations,
\begin{equation}
 \rho_\phi 
  = \rho_{\phi\, \mathrm{q}} \, 
  \frac{s}{s_{\mathrm{q}}}
   \simeq \frac{1}{2} m^2 \lambda^2 \frac{s}{s_{\mathrm{osc}}}
 \frac{s_{\mathrm{osc}}}{s_{\mathrm{q}}}
  = \frac{1}{2} \frac{s}{s_{\mathrm{osc}}}
  \frac{m^2 \phi_{\mathrm{osc}}^{5/2}}{\lambda^{1/2}}
  \qquad
  \mathrm{for}
  \quad
  t \geq t_{\mathrm{q}}.
\label{phi-abund}
\end{equation}
In a RD universe, the entropy density can be expressed in terms of
the Hubble rate using (\ref{entden}) and (\ref{RDuniv}).
Further using (\ref{Hosc}) and (\ref{phi_osc}),
the energy density of the oscillating scalar
while the universe is RD
can also be expressed as
\begin{equation}
 \rho_{\phi} \simeq
  \frac{1}{2} \left(\frac{10}{3}\right)^{3/4}
  \left(\frac{6}{7}\right)^{13/4}
  \frac{g_{s*}}{g_{s*\mathrm{osc}}}
  \left( \frac{g_{*\mathrm{osc}}}{g_*} \right)^{3/4}
  \frac{H^{3/2} m^{1/2} \phi_{\star}^{13/4} }{\lambda^{5/4}}
    \qquad
  \mathrm{for}
  \quad
  t_{\mathrm{q}} \leq t <  t_{\mathrm{eq}} ,
\label{phi-abund-H}
\end{equation}
where $t_{\mathrm{eq}}$ denotes the time of matter-radiation
equality.\footnote{The anharmonic oscillations along the potential $V
\propto \sigma^{4/3}$, which is flatter than a quadratic, may lead to
the formation of localized configurations of the oscillating scalar
field~\cite{Bogolyubsky:1976nx,Gleiser:1993pt,Copeland:1995fq}.
Here we assumed that such oscillons do not form, but if they do,
the final scalar density could be modified from~(\ref{phi-abund-H}).}

\subsubsection*{Decay of Scalar}

After the harmonic oscillation begins, the scalar eventually 
decays away through the derivative coupling~$(\partial_\mu \phi) j^\mu / f$.
The decay modes depend on the particles constituting the 
current~$j^\mu$, however the derivative coupling typically includes
anomalous couplings to $W\tilde{W}$ and $Z\tilde{Z}$.
Here we do not specify the decay channel, and instead parameterize the decay
rate of the scalar by
\begin{equation}
 \Gamma_\phi = \frac{\beta }{64 \pi^3} \frac{m^3}{f^2},
\label{Gamma_n1}
\end{equation}
with a dimensionless constant~$\beta$  which is typically smaller than unity.

The scalar decays when the Hubble expansion rate becomes
comparable to the decay rate, i.e., around when $H = \Gamma_\phi$.
If this happens during $t_{\mathrm{q}} \leq t <  t_{\mathrm{eq}}$,
then the ratio between the scalar energy density and the total density of
the universe right before the decay is estimated
from~(\ref{phi-abund-H}) as
\begin{equation} 
 r \equiv
  \left. \frac{\rho_{\phi }}{3 M_p^2 H^2 } \right|_{H = \Gamma_\phi} = 
 \frac{2^6 \cdot 3^{3/2} \cdot 5^{3/4} \pi^{3/2}}{7^{13/4} \beta^{1/2}}
  \frac{g_{s* \mathrm{D}}}{g_{s*\mathrm{osc}}}
  \left( \frac{g_{*\mathrm{osc}}}{g_{*\mathrm{D}}} \right)^{3/4}
  \frac{ f \phi_{\star}^{13/4} }{  M_p^2 m  \lambda^{5/4} },
\label{2.30}
\end{equation}
where $g_{(s)* \mathrm{D}}$ corresponds to the effective relativistic
degrees of freedom right before the decay. 
Upon obtaining this expression we have considered a RD universe, 
hence the oscillating scalar (which behaves as pressureless dust) is
assumed to be subdominant, i.e. $ r < 1$.
On the other hand if $r$~as expressed in~(\ref{2.30}) exceeds unity,
then it would signify that the scalar dominates the universe
before decaying away.
However in such cases, the entropy production by the decay of the
dominant scalar would dilute the already produced baryon
asymmetry.\footnote{A possibility that we do not pursue in
this work is that a baryon asymmetry much larger than in the present
universe is originally produced, but gets diluted by the scalar
domination. Although the parameter space for such scenarios is expected
to be quite narrow,
as the scalar can impact the cosmological expansion history by once dominating
the universe, it would be interesting to investigate such cases with an
eye towards the observational signals the scalar may leave.}

Here one may wonder whether the scalar could serve as dark matter if it
survives until the present;
however one can show that a stable scalar would overclose the universe.
Neglecting for the moment the decay,
the scalar abundance today is computed by substituting the
present day entropy density~$s_0$ into (\ref{phi-abund}),
which can be expressed as
\begin{equation}
\begin{split}
 \Omega_\phi h^2 \equiv \frac{\rho_{\phi 0} h^2}{3 M_p^2 H_0^2}
 \approx 300
    & \times
  \Bigl(\sum_i B_i c_i g_i \Bigr)^{-3}
  \left( \frac{g_{*\mathrm{dec}}}{106.75}  \right)^{-5/4}
  \left( \frac{g_{s*\mathrm{dec}}}{106.75}  \right)^{3}
  \left( \frac{g_{*\mathrm{osc}}}{106.75}  \right)^{3/4}
  \left( \frac{g_{s*\mathrm{osc}}}{106.75}  \right)^{-1}
 \\ & \times
  \left( \frac{(n_B/s)_0}{8.6 \times 10^{-11}} \right)^3
  \left( \frac{f  }{\lambda} \right)
  \left( \frac{\phi_{\star}}{\lambda } \right)^{1/2}
  \left( \frac{H_{\mathrm{dec}}}{H_{\mathrm{osc}}} \right)^{11/2}
  \left( \frac{f}{T_{\mathrm{dec}}} \right)^2.
\label{2.31}
\end{split}
\end{equation}
Here we have used (\ref{2.21}) as the baryon-to-entropy ratio today,
and also (\ref{RDuniv}), (\ref{cond-ii}). 
The dimensionless Hubble constant~$h$ is defined as
$H_0 = 100 h \, \mathrm{km} \, \mathrm{sec}^{-1} \, \mathrm{Mpc}^{-1}$.
Each of the last four parentheses in the second line is larger than
unity, as we have been assuming 
the initial field value to lie within the range of $\lambda \ll
\phi_\star \lesssim f$, 
the onset of oscillations to be after decoupling, i.e. 
$H_{\mathrm{osc}} < H_{\mathrm{dec}}$,
and the decoupling temperature to satisfy $T_{\mathrm{dec}} < f$
(cf. discussions below~(\ref{Sn1})).
Hence as long as $\sum_i B_i c_i g_i \sim 1$
and $g_{(s)*} \sim 100$ at the relevant times,
then the (hypothetical) relic abundance of a scalar that creates the
observed baryon asymmetry would be as large as
$\Omega_\phi h^2 \gtrsim 300$.
This indicates that, were it not for the decay,
the scalar would dominate the universe well before the standard
matter-radiation equality.

\subsection{Constraint on Curvaton-like Behaviors}
\label{subsec:curvaton-like}

It should also be noted that 
if the scalar dominates or comes close to dominating the universe before
decaying, it would create curvature perturbations {\it \` a la}
curvatons~\cite{Linde:1996gt,Enqvist:2001zp,Lyth:2001nq,Moroi:2001ct}. 
In such cases,
although our model blocks baryon isocurvature during baryogenesis, 
the creation of the curvature perturbations in the end gives rise to 
baryon isocurvature. 
Here we estimate this effect and place an upper bound on
the density ratio~$r$ upon decay~(\ref{2.30})
as well as the inflation scale.

From (\ref{phi-abund-H}) we see that the scalar density at decay depends
on the scalar field value during inflation as\footnote{Strictly speaking,
$g_{(s)*\mathrm{osc}}$ can also depend on~$\phi_\star$, but we ignore this
effect here.}
$\rho_\phi |_{H = \Gamma_\phi} \propto \phi_\star^{13/4} \propto
\sigma_{\star}^{13/6}$.  
Thus the curvature perturbation produced by the scalar is estimated,
up to linear order in the field fluctuation, as
\begin{equation}
 \zeta_{\phi} \sim \min (r,1)
  \left. 
   \frac{\delta \rho_{\phi}}{\rho_{\phi}}
   \right|_{H = \Gamma_\phi}
  = \min (r,1)\,  \frac{13}{6}
  \frac{\delta \sigma_\star}{\sigma_\star}.
\label{zeta_phi}
\end{equation}
During inflation the field~$\sigma$ is nearly canonical
with an effective mass
much lighter than the Hubble rate, therefore 
its fluctuation power spectrum is
$\mathcal{P}_{\delta \sigma_\star}  = (H_{\mathrm{inf}} / 2 \pi
)^2$.
The curvature perturbation created after the production of the
baryon asymmetry sources a baryon isocurvature of
$S_{B \gamma} = -3 \zeta_{\phi}$
(see e.g. Eq.~(5.6) of~\cite{DeSimone:2016ofp}), hence 
the resulting baryon isocurvature power on the pivot scale is
\begin{equation}
 \mathcal{P}_{B \gamma} (k_\star)
  \sim
  \left\{
\min (r,1)\,  \frac{13}{2} \frac{H_{\mathrm{inf}}}{2 \pi \sigma_\star}
  \right\}^2
\lesssim 3 \times 10^{-9}.
\end{equation}
In the far right hand side,
we have shown the current upper limit on a scale-invariant and uncorrelated
baryon isocurvature perturbation from the {\it Planck} CMB results~\cite{Ade:2015lrj}.
Thus we obtain an order-of-magnitude constraint on the curvaton-like
behavior of the scalar field as
\begin{equation}
 \min (r,1)\,  \frac{H_{\mathrm{inf}}}{ \sigma_\star}
  \lesssim 10^{-4}.
  \label{curvaton-bound}
\end{equation}
One can also require that the curvature
perturbation~(\ref{zeta_phi}) produced by the scalar should not exceed
the observed amplitude $\mathcal{P}_\zeta (k_\star) \sim 2 \times
10^{-9}$, however this gives a bound weaker
than~(\ref{curvaton-bound}).

\subsection{Parameter Space}

We now look into the parameter space of our model that allows a
successful baryogenesis.
Let us recall the requirements.

First of all, we have considered a cosmological history where inflation is followed by
reheating, and then by the decoupling of the baryon violating
interactions,
i.e. $H_{\mathrm{inf}} > H_{\mathrm{reh}} > H_{\mathrm{dec}}$.
The baryon isocurvature can be suppressed if the
scalar is initially located in the large field
regime of $\lambda \ll \phi_{\star} \lesssim f$.
Here, the scale of~$f$ is assumed to be higher than the inflationary
Hubble rate as well as the reheating temperature, i.e.
$f > H_{\mathrm{inf}}, T_{\mathrm{reh}}$.
This in particular implies $f > T_{\mathrm{dec}}$.
Moreover, the initial value of the canonical field is required to be
larger than the field 
fluctuations obtained during inflation,
i.e. $\sigma_\star > H_{\mathrm{inf}}$ (cf.~(\ref{super-Hinf})), 
but sub-Planckian,
i.e. $\sigma_\star < M_p$ (cf.~(\ref{sub-Planck})), to avoid the
scalar itself from driving inflation or dominating the universe.
Then the baryon asymmetry is generated as~(\ref{2.21}),
given that the scalar initially obeys the slow-varying
conditions~(\ref{SVcond}),
and also that the baryon violating interactions decouple
before the scalar starts its oscillations,
i.e. $H_{\mathrm{dec}} >  H_{\mathrm{osc}}$ (cf.~(\ref{cond-ii})).
Upon computing $\phi_{\mathrm{osc}}$ 
we have also assumed a hierarchy between the Hubble rates as
$H_{\mathrm{osc}}^2 \ll H_{\mathrm{reh}}^2$ and 
$H_{\mathrm{osc}}^2 \ll  3 H_{\mathrm{inf}}^2 / (20 \mathcal{N}_\star)$ 
(cf.~(\ref{2.26}));
the latter condition implies that the scalar field value is frozen during inflation.

Here, it is easy to check that the three slow-varying
conditions of~(\ref{SVcond}) follow from the other requirements:
The first condition is nothing but the 
large field requirement.
The second condition is guaranteed to hold at least until around
decoupling from $H_{\mathrm{dec}} > H_{\mathrm{osc}}$.
As for the third condition which is required during baryogenesis,
we have already seen in~(\ref{HAinT}) that
it is automatically satisfied for $\lambda \ll \phi_\star$ and $f > T$.

In order to avoid the baryon asymmetry from being diluted
after the scalar starts to oscillate,
the scalar is required to decay away before dominating the
universe, i.e. $r < 1$, where the density ratio~$r$
at decay is given in~(\ref{2.30}).
Here the scalar is considered to decay during its harmonic oscillations, thus
we have assumed its decay rate to satisfy
$\Gamma_{\phi} < H_{\mathrm{q}}$,
where $H_{\mathrm{q}}$ can be obtained from~(\ref{sqsosc}).
The condition $ r < 1$ guarantees the scalar to decay before the
matter-radiation equality, but in order for the scalar not to spoil Big
Bang Nucleosynthesis (BBN), we further require the scalar to either
decay before BBN, or have negligible density during the period. 
The ratio~$r$, combined with the inflation scale, is further
bounded as~(\ref{curvaton-bound}) from the
constraint on the curvaton-like behavior of the scalar. 

\vspace{\baselineskip}

\begin{figure}[t]
\centering
\subfigure[$\lambda = 10^{-1} f$, $\phi_\star = f$]{%
  \includegraphics[width=.475\linewidth]{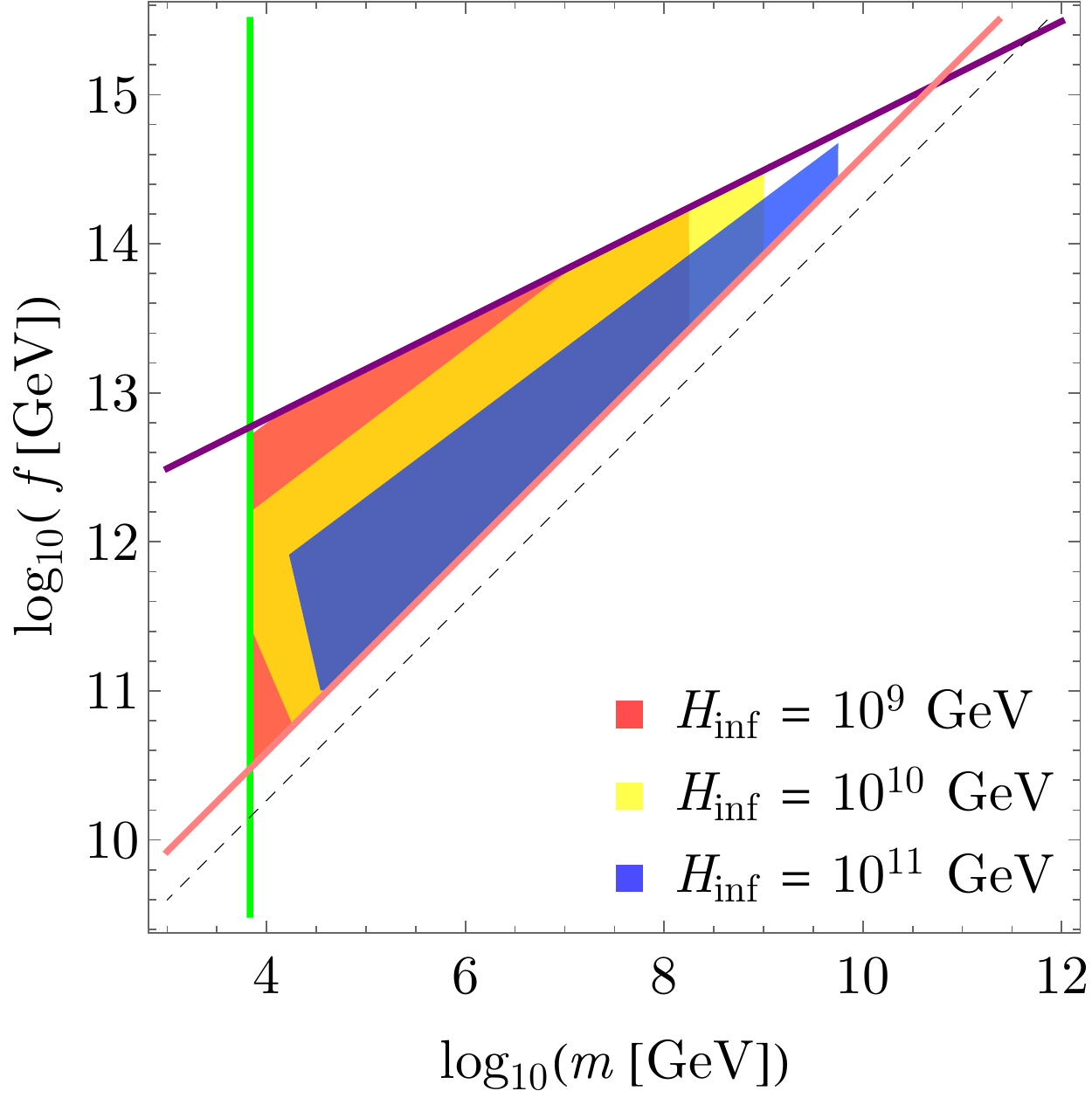}
  \label{fig:frac1}}
\, 
\subfigure[$\lambda = 10^{-4} f$, $\phi_\star = 10^{-2} f$]{%
  \includegraphics[width=.475\linewidth]{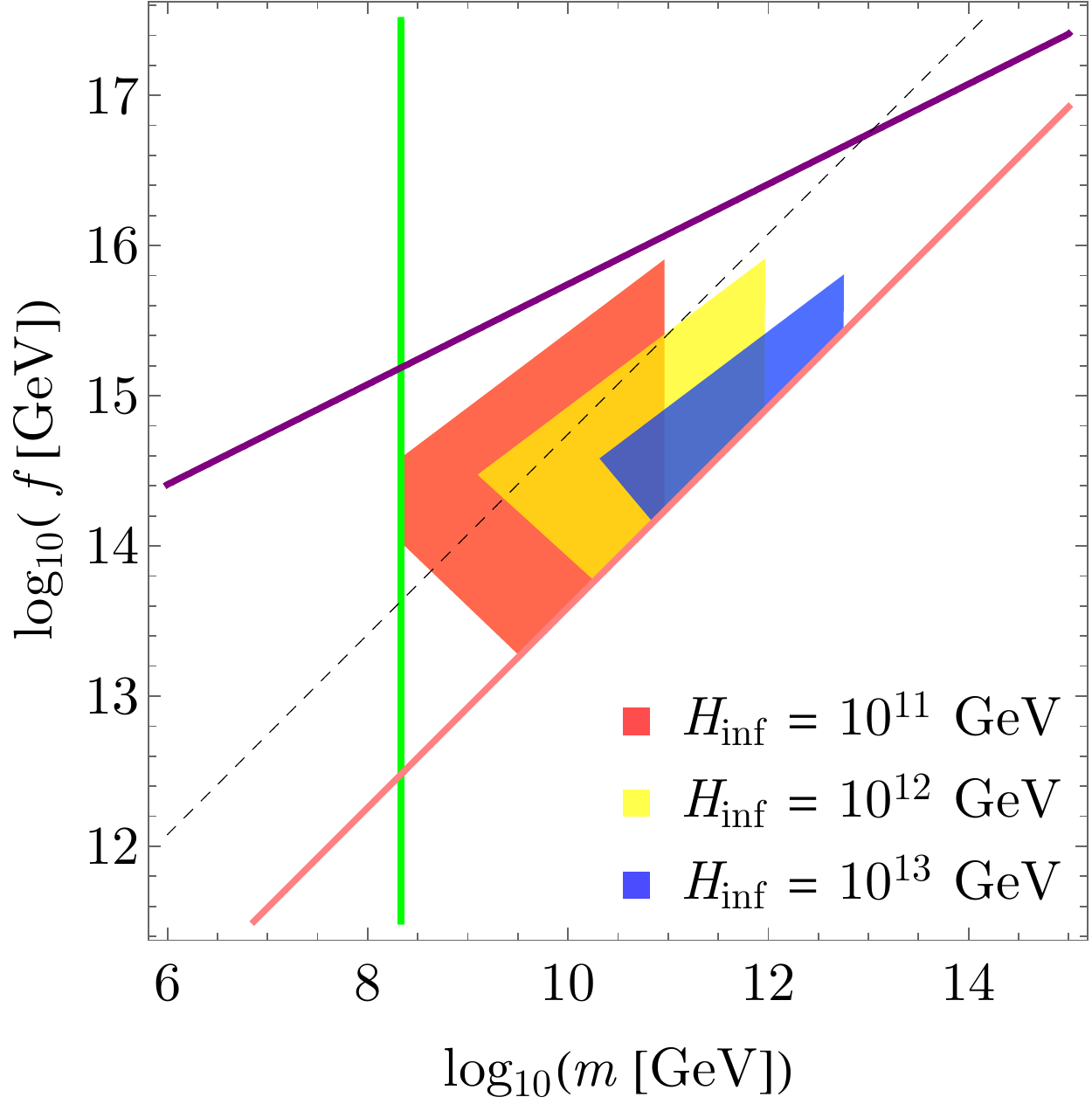}
  \label{fig:frac2}}
 \caption{Parameter space for the $n = 1$ spontaneous baryogenesis model in the $m$ -- $f$ plane.
 The viable parameter space is shown by the colored regions, where
 the different colors correspond to different inflation scales. 
 The solid lines represent the requirements for the model: decoupling
 before the onset of scalar oscillations $H_{\mathrm{dec}} >
 H_{\mathrm{osc}}$ (green), decay of the scalar before dominating the
 universe $r < 1$ (purple), and $ f > T_{\mathrm{dec}} $ (pink).
 The black dashed lines indicate where the model may become sensitive to
 the UV theory.
 The other conditions are explained in the text.}
\label{fig:frac}
\end{figure}

In Figure~\ref{fig:frac} we show the viable parameter space in the $m$ -- $f$ plane. 
Here the dimensionless parameters are chosen as
$\sum_{i} B_i c_i g_i = 1$, $\beta = 1$, and
the relativistic degrees of freedom is considered to be
$ g_{(s)*} = 106.75$ 
from the time of decoupling until the decay of the scalar. 
We have also fixed the decoupling temperature~$T_{\mathrm{dec}}$
from the requirement that the final baryon-to-entropy ratio~(\ref{2.21})
matches the present value of $(n_B / s)_0 \approx 8.6 \times 10^{-11}$. 
Furthermore, the values of $\lambda$ and $\phi_\star$ are fixed in terms
of~$f$ so that they satisfy $\lambda \ll \phi_\star \lesssim f$;
we show example cases where
$\lambda = 10^{-1} f$, $\phi_\star = f$ (left panel) and 
$\lambda = 10^{-4} f$, $\phi_\star = 10^{-2} f$ (right panel).
The solid lines in the figures represent the conditions that do not
involve the inflation scale~$H_{\mathrm{inf}}$, which are
the requirements of decoupling before the onset of the scalar oscillations 
$H_{\mathrm{dec}} > H_{\mathrm{osc}}$ (green line),
the scalar decay before dominating the universe 
$r < 1$ (purple line),
and $ f > T_{\mathrm{dec}}$ (pink line).
The triangular regions surrounded by these solid lines satisfy
all three requirements.
We also note that in the entire triangular regions, 
the field value is sub-Planckian $\sigma_\star < M_p$,
and the decay rate of the scalar satisfies 
$H_{\mathrm{BBN}} < \Gamma_\phi < H_{\mathrm{q}} $
so that the scalar decays after starting its harmonic oscillations, but
before~BBN.

The allowed windows further shrink when taking into account the
conditions that involve the inflation scale~$H_{\mathrm{inf}}$.
The colored regions show the parameter space satisfying all conditions,
under fixed values of~$H_{\mathrm{inf}}$.
Here we have also numerically computed the baryon isocurvature
perturbations that arise due to deviations from the slow-varying
solution, which was discussed below~(\ref{2.21}).
We numerically solved the scalar's equation of motion
starting from initial conditions of~$\sigma_\star$ that vary
by~$H_{\mathrm{inf}} / 2\pi$,
and estimated the baryon isocurvature by computing the difference 
arising in the chemical potential $\mu \propto \dot{\phi}$ at the time of
decoupling.\footnote{We remark that the value of the reheating
temperature, as long as it satisfies the conditions mentioned above, has
little effect on the amplitude of the baryon isocurvature. We also note
that we have neglected the backreaction from the baryons in the
numerical computations.} In corners of the parameter space where some
conditions are only marginally satisfied
(such as cases where the slow-varying conditions~(\ref{SVcond}) start to
break down at decoupling),
the baryon isocurvature can become non-negligible, as we see below.

In the left panel for $\lambda = 10^{-1} f$ and $\phi_\star = f$,
we show the viable parameter space under
$H_{\mathrm{inf}} = 10^9\, \mathrm{GeV}$ (red
shaded region), $10^{10}\, \mathrm{GeV}$ (yellow shaded region), and
$10^{11}\, \mathrm{GeV}$ (blue shaded region).
Note that the regions overlap with each other;
in particular the left part of the blue region
is on top of the yellow region, whose left part is on top of the red region.
Thus the red region actually occupies
most of the triangular region surrounded by the solid lines.
The right-side boundaries of each colored regions are determined by the 
condition of $ H_{\mathrm{inf}} > H_{\mathrm{dec}}$,
while the upper left boundaries are from the
constraint~(\ref{curvaton-bound}) on the curvaton-like behavior.
In the lower left corner of the triangular region,
the ratio $H_{\mathrm{inf}} / \sigma_\star$ is not too small,
and moreover the condition $H_\mathrm{dec} > H_{\mathrm{osc}}$ is only
marginally satisfied.\footnote{If $H_{\mathrm{osc}}$ is close
to~$H_{\mathrm{dec}}$, the rapid rolling of the scalar towards the end
of baryogenesis may also lead to a large time-variation of the chemical
potential~$\mu$. It was discussed in~\cite{Arbuzova:2016qfh} that in
such cases the computation of the baryon production can be modified.} 
This implies that the scalar at the time of
decoupling is starting to deviate from the slow-varying solution,
and hence a non-negligible fraction of 
the large field fluctuations can leak into the baryon isocurvature. 
Comparison of the numerically computed baryon isocurvature with the {\it
Planck} bound trims off the lower left corners of the windows for
$H_{\mathrm{inf}} = 10^{10}\, \mathrm{GeV}$ and $10^{11}\, \mathrm{GeV}$,
as shown in the figure.

In the right panel for $\lambda = 10^{-4} f$ and $\phi_\star = 10^{-2} f$,
we show the viable parameter space for 
$H_{\mathrm{inf}} = 10^{11}\, \mathrm{GeV}$ (red
shaded region), $10^{12}\, \mathrm{GeV}$ (yellow shaded region), and
$10^{13}\, \mathrm{GeV}$ (blue shaded region).
The right-side boundaries of the windows for
$H_{\mathrm{inf}} = 10^{11},\, 10^{12}\, \mathrm{GeV}$
are from the condition $H_{\mathrm{osc}}^2 <  3 H_{\mathrm{inf}}^2 / (20
\mathcal{N}_\star)$, 
while the right-side boundary for $H_{\mathrm{inf}} = 10^{13}\,
\mathrm{GeV}$ is from $ H_{\mathrm{inf}} > H_{\mathrm{dec}}$
(which of these conditions is stronger depends on the inflation scale).
The upper left boundaries for all three cases are from the 
constraint~(\ref{curvaton-bound}) on the curvaton-like behavior,
and the lower left boundaries are from the constraint on the baryon
isocurvature arising from deviations from the slow-varying trajectory. 
The parameter window here exists for inflation scales up to 
$H_{\mathrm{inf}} \sim 10^{14}\, \mathrm{GeV}$,
which implies that spontaneous baryogenesis can be compatible even with
high inflation scales close to the current observational upper
limit~\cite{Ade:2015lrj}. 

Note also that each point in the colored regions possesses a range for
the reheating scale that satisfies
$  H_{\mathrm{inf}} > H_{\mathrm{reh}}$ and 
$f > T_{\mathrm{reh}} > T_{\mathrm{dec}}$.
However, we should also remark that the allowed range
of~$T_{\mathrm{reh}}$ becomes small in regions 
close to the boundaries where the
conditions such as 
$ H_{\mathrm{inf}} > H_{\mathrm{dec}}$
or $ f > T_{\mathrm{dec}} $ 
are only marginally satisfied.
In such corners of the parameter space, one will have to require an
instantaneous reheating and/or the baryon violating processes to
decouple soon after reheating.

We further remark that the model may become sensitive to the UV completion
of the theory in some parts of the parameter space.
Such regions are indicated by the black dashed lines in the figures,
which we will discuss in the next section.

\subsection{Power Counting Estimate of Cutoff}
\label{subsec:cutoff}

Although the main focus of the present paper is on the phenomenology of
spontaneous baryogenesis with a $\phi$-independent chemical
potential,
let us briefly comment on the validity of the 
theory~(\ref{Sn1}) by estimating its cutoff scale, above which the
effective field theory breaks down and new physics should intervene.
Here we estimate a lower bound on the cutoff through power counting
analyses as was discussed
in, e.g.,~\cite{Burgess:2009ea,Hertzberg:2010dc,Bezrukov:2010jz}.  
The cutoff depends on the background scalar value, hence let us focus on
the large field regime $\sigma \gg \lambda$ 
during spontaneous baryogenesis,
and expand the almost canonical~$\sigma$ around its background
value as
$\sigma = \sigma_{\mathrm{bg}} + \hat{\sigma}$.
Then the scalar potential~(\ref{V4over3}) yields operators involving
$p$~powers of $\hat{\sigma}$ as
\begin{equation}
 m^2 \lambda^{\frac{2}{3}} \frac{\hat{ \sigma}^p}{\sigma_{\mathrm{bg}}^{
  p - \frac{4}{3}} }, 
\label{op1}
\end{equation}
where we ignored numerical coefficients.
Hence from operators with dimension~$p$ larger than four,
the energy scales above which perturbation theory breaks down can be read off as
\begin{equation}
 \Lambda_{1(p)} \sim
  \left(\frac{\sigma_{\mathrm{bg}}^{p-\frac{4}{3}}}{m^2
   \lambda^{\frac{2}{3}}}\right)^{\frac{1}{p-4}}.
\end{equation}
Likewise, the derivative coupling to the baryon current
in~(\ref{eq3.3}), after integration by parts, yields operators with $ p
\geq 5$ of the form
\begin{equation}
 \frac{\lambda^{\frac{1}{3}}}{f}
   \frac{\hat{\sigma}^{p-4}}{\sigma_{\mathrm{bg}}^{p-\frac{14}{3}}}
  \sum_{i} c_i \nabla_\mu j_i^{\mu},
\label{op2}
\end{equation}
with which perturbative analyses are expected to break down above
\begin{equation}
 \Lambda_{2(p)} \sim \left( \frac{f}{ \lambda^{\frac{1}{3}}}
	\sigma_{\mathrm{bg}}^{p-\frac{14}{3}}  \right)^{\frac{1}{p-4}}.
\end{equation}
These estimates imply the cutoff of the theory as
$\Lambda = \min_{p \geq 5} (\Lambda_{1(p)}, \Lambda_{2(p)})$.
Here, note that both $\Lambda_{1(p)}$ and $\Lambda_{2(p)}$ asymptote
to~$\sigma_{\mathrm{bg}}$ in the large $p$ limit.

Representing the typical energy scale during baryogenesis
by the decoupling temperature, let us now compare it to~$\Lambda$.
In Figure~\ref{fig:frac},
by taking $\sigma_{\mathrm{bg}} = \sigma_\star$, we have indicated where
$T_{\mathrm{dec}} = \Lambda$ by the black dashed lines;
$T_{\mathrm{dec}} < \Lambda$ is satisfied on the left sides of the lines.
For the chosen sets of parameters,
and further in the regions of $ m < f$ where the allowed windows exist,
$\Lambda_{1,2(p)}$ are either independent of~$p$, or become smaller for
larger~$p$.
Hence $\Lambda = \Lambda_{1,2(p \to \infty)} \sim \sigma_{\mathrm{bg}}$.
In Figure~\ref{fig:frac1}, one sees that $T_{\mathrm{dec}}$ is safely
below~$\Lambda$ in the entire allowed window.
On the other hand in Figure~\ref{fig:frac2},
$T_{\mathrm{dec}}$ exceeds $\Lambda$ in some part of the window;
there the model may be sensitive to the UV completion of the theory.
One can further estimate the cutoff in the small field regime
$\abs{\phi} \ll \lambda$ during the scalar oscillations and require it
to be larger than the oscillation energy $\rho_\phi^{1/4} \sim m^{1/2}
\phi^{1/2}$,
but this turns out to yield weaker constraints on the allowed windows
compared to those during baryogenesis.

However, we also remark that the constraint of $T_{\mathrm{dec}} <
\Lambda$ should be considered as a rather conservative bound,
as the typical energy scales of the scattering processes described by
the operators (\ref{op1}) and (\ref{op2}) may actually be lower than
$T_{\mathrm{dec}}$. 
Furthermore, provided the theory is endowed with some extra symmetries,
the naive power counting can fail and the cutoff can be much higher than
the $\Lambda$ estimated above. 
We leave a more detailed analysis of these issues, including explicit
computations of the scattering amplitudes, for future work.

\section{$n = 2$ : Spontaneous Baryogenesis with Linear Terms}
\label{sec:lin}

In this section we study spontaneous baryogenesis driven by a scalar with
a $Z_2$~symmetric action of
\begin{equation}
 S = \int d^4 x \sqrt{-g}
  \left[
   -\frac{1}{2} \left\{ 1 + \left(\frac{\phi }{\lambda }\right)^{2} \right\}
   g^{\mu  \nu}  \partial_\mu \phi \partial_\nu \phi
   - \frac{1}{2} m^2 \phi^2
   - \frac{1}{2} \left( \frac{\phi }{f} \right)^2 \sum_i c_i
   \nabla_\mu j_{i}^\mu
  \right],
\label{Sn2}
\end{equation}
which is phenomenologically similar to the case of $n = 2$ in (\ref{Ln}).
The field that becomes canonical in the large field regime of $\phi \gg
\lambda$ is 
\begin{equation}
 \sigma = \frac{1}{2} \frac{\phi^{2}}{\lambda},
\end{equation}
with which the action is rewritten as, after integrating by parts the
coupling term, 
\begin{equation}
 S = \int d^4 x \sqrt{-g}
  \Biggl[
 -\frac{1}{2}
\left( 1 + \frac{\lambda }{2 \sigma } \right)
 g^{\mu \nu} \partial_\mu \sigma
\partial_\nu \sigma
- m^2 \lambda \sigma 
 + \frac{\lambda \, \partial_\mu \sigma }{f^2 }
\sum_i c_i j_i^{\mu}
  \Biggr].
\end{equation}
Hence this model in the large field regime reduces to a theory with a
linear potential as well as a linear coupling to the baryon current.

The basic picture is the same as for the $n = 1$ case studied in the
previous section, hence here we just list the relevant results that
can be obtained from similar computations.
The final baryon-to-entropy ratio is now
\begin{equation}
 \left. \frac{n_B}{s} \right|_{\mathrm{dec}} =
    \frac{9}{\pi^3} \left(\frac{5}{8}\right)^{1/2}
  \frac{\sum_i B_i c_i g_i }{
g_{*\mathrm{dec}}^{1/2} \,  g_{s*\mathrm{dec}} }
\frac{M_p m^2 \lambda^2 }{T_{\mathrm{dec}}^3 \,  f^2},
  \label{3.4}
\end{equation}
and one can also check that the ratio between the Hubble rates at 
the onset of the oscillations
and decoupling, which should be smaller than unity, is
\begin{equation}
 \left(
\frac{H_{\mathrm{osc}}}{H_{\mathrm{dec}}}
 \right)^2
 = \frac{45 }{\pi^2 } \frac{1}{g_{*\mathrm{dec}}}
 \frac{M_p^2 m^2 \lambda^2 }{T_{\mathrm{dec}}^4 \phi_{\star}^2 } < 1.
\label{cond-ii-n2}
\end{equation}
During the anharmonic oscillation along the linear potential,
the scalar density and oscillation amplitude redshift as
\begin{equation}
 \rho_\phi \propto a^{-2},
  \quad
   \bar{\sigma} \propto a^{-2},
    \quad
  \bar{\phi} \propto a^{-1},
\end{equation}
and the scalar density after the harmonic oscillation begins is
obtained as
\begin{equation}
 \rho_{\phi} \simeq
\frac{2^{11/4}}{5^{3/2}}
  \frac{g_{s*}}{g_{s*\mathrm{osc}}}
  \left( \frac{g_{*\mathrm{osc}}}{g_*} \right)^{3/4}
  \frac{H^{3/2} m^{1/2} \phi_{\star}^{9/2} }{\lambda^{5/2}}
    \qquad
  \mathrm{for}
  \quad
  t_{\mathrm{q}} \leq t <  t_{\mathrm{eq}}.
\label{phi-abund-H-n2}
\end{equation}
As in the case of $n = 1$, were it not for the decay, the oscillating
scalar would overclose the universe well before the standard
matter-radiation equality. 
However, since the Lagrangian for~$\phi$ is now $Z_2$~symmetric, the scalar
condensate could be stable and thus be disastrous for
cosmology.\footnote{Although, depending on the parameters, the 
scalar~$\phi$ may dissipate its energy through the $\phi^2 \nabla_\mu
j^\mu$ interaction. For detailed discussions on $Z_2$ symmetric scalars,
see e.g. \cite{Mukaida:2013xxa} and references therein.}

Hence let us suppose that the $Z_2$ symmetry is slightly broken so that
the minima of the quadratic potential and the derivative coupling term
are misaligned by $\Delta \phi \sim \lambda$. 
Then, expanding the coupling term around the potential minimum gives
rise to
$ \sim (\lambda \phi / f^2) \sum_i c_i \nabla_\mu j_{i}^\mu$,
providing a decay channel for~$\phi$.
Expressing the decay rate by
\begin{equation}
 \Gamma_\phi = \frac{\beta }{64 \pi^3} \frac{m^3 \lambda^2}{f^4},
\label{Gamma_n2}
\end{equation}
with a dimensionless parameter~$\beta$ which is typically smaller than
unity, one can compute the 
density ratio right before the decay of the scalar as
\begin{equation} 
 r \equiv
  \left. \frac{\rho_{\phi }}{3 M_p^2 H^2 } \right|_{H = \Gamma_\phi} = 
  \frac{2^{23/4} \pi^{3/2}}{3 \cdot 5^{3/2} \beta^{1/2}}
  \frac{g_{s* \mathrm{D}}}{g_{s*\mathrm{osc}}}
  \left( \frac{g_{*\mathrm{osc}}}{g_{*\mathrm{D}}} \right)^{3/4}
  \frac{ f^2 \phi_{\star}^{9/2} }{  M_p^2 m  \lambda^{7/2} },
  \label{r_for_n2}
\end{equation}
given that the scalar decays during $ t_{\mathrm{q}} \leq t <  t_{\mathrm{eq}}$.

\begin{figure}[t]
\centering
\subfigure[$\lambda = 10^{-1} f$, $\phi_\star = f$]{%
  \includegraphics[width=.475\linewidth]{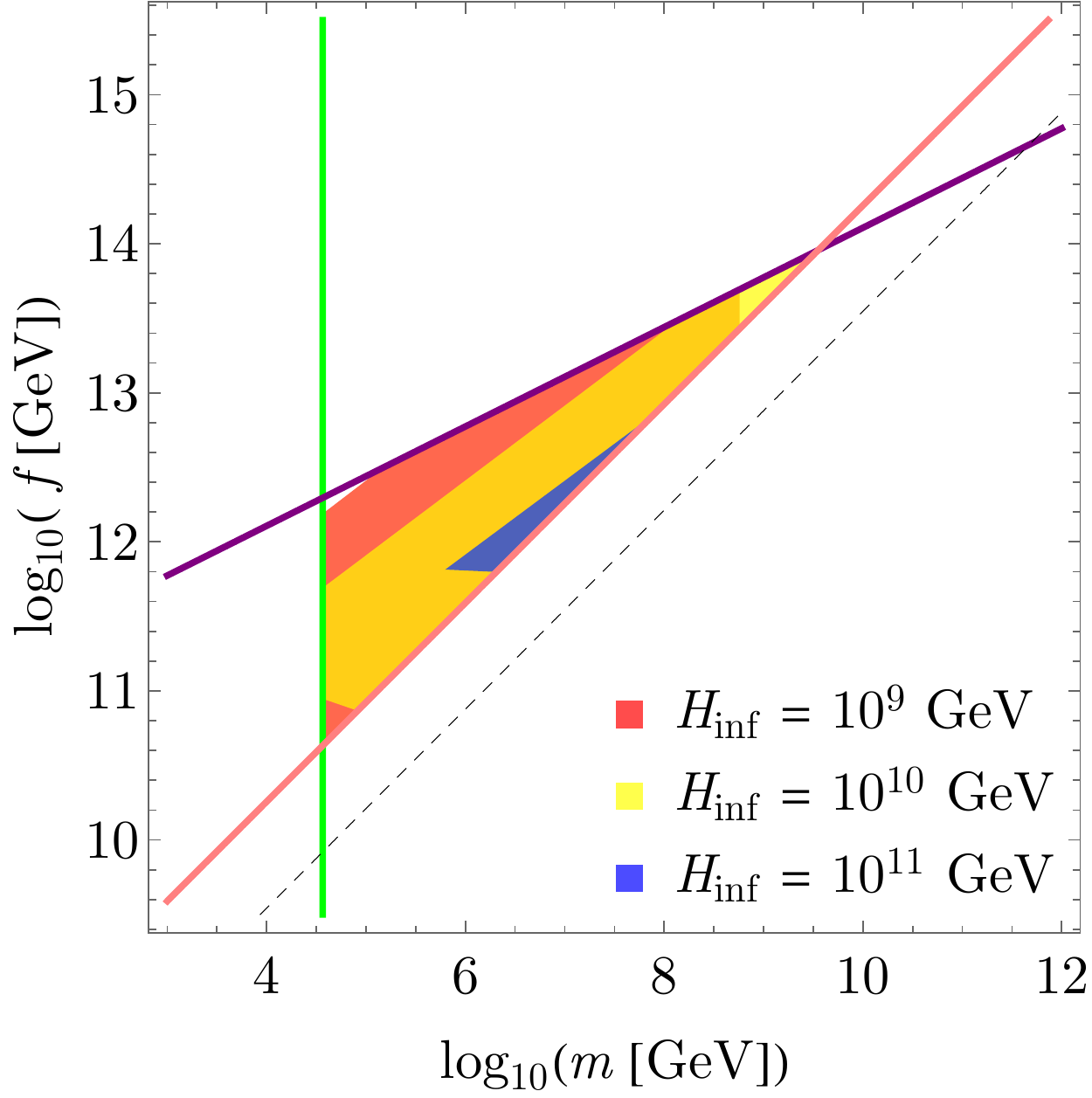}
  \label{fig:lin1}}
 \caption{Parameter space for the $n = 2$ spontaneous baryogenesis model in the $m$ -- $f$ plane.
 The viable parameter space is shown by the colored regions, where
 the different colors correspond to different inflation scales. 
 The solid lines represent the requirements for the model: decoupling
 before the onset of scalar oscillations $H_{\mathrm{dec}} >
 H_{\mathrm{osc}}$ (green), decay of the scalar before dominating the
 universe $r < 1$ (purple), and $  f > T_{\mathrm{dec}} $ (pink).
 The black dashed line indicates where the model may become sensitive to
 the UV theory.
 The other conditions are explained in the text.}
\label{fig:lin}
\end{figure}

The baryon-to-entropy ratio~(\ref{3.4}) and the decay
rate~(\ref{Gamma_n2}) in the $n = 2$  model 
are suppressed by powers of $(\lambda / f)$
compared to the case of $n = 1$, cf. (\ref{2.21}) and (\ref{Gamma_n1}).
As a consequence, the parameter space is narrower for $n = 2$. 
The $m$ -- $f$ parameter space for the $n = 2$ model is shown in
Figure~\ref{fig:lin}.
The dimensionless parameters are chosen as
$\sum_{i} B_i c_i g_i = 1$, $\beta = 1$, $ g_{(s)*} = 106.75$,
and the decoupling temperature~$T_{\mathrm{dec}}$ is fixed
by normalizing the baryon-to-entropy ratio~(\ref{3.4}) to the
present value $(n_B / s)_0 \approx 8.6 \times 10^{-11}$.
Furthermore, $\lambda$ and $\phi_\star$ are fixed as 
$\lambda = 10^{-1} f$, $\phi_\star = f$.
The conditions that do not involve~$H_{\mathrm{inf}}$ are represented by
the solid lines:
$H_{\mathrm{dec}} > H_{\mathrm{osc}}$ (green line),
$r < 1$ (purple line),
and $ f >  T_{\mathrm{dec}} $ (pink line).
The triangular region surrounded by these solid lines satisfy
the three requirements, as well as
$\sigma_\star < M_p$ and $H_{\mathrm{BBN}} < \Gamma_\phi <
H_{\mathrm{q}} $.
After further taking into account the conditions on~$H_{\mathrm{inf}}$,
then the viable parameter space satisfying all conditions are shown as the
colored regions. 
In each colored region, the inflation scale is fixed to
$H_{\mathrm{inf}} = 10^9\, \mathrm{GeV}$ (red shaded region),
$10^{10}\, \mathrm{GeV}$ (yellow shaded region),
$10^{11}\, \mathrm{GeV}$ (blue shaded region).
Here, the constraints that set further boundaries
inside the triangular region are:
$H_{\mathrm{inf}} >  H_{\mathrm{dec}}$ (right-side boundary for
$H_{\mathrm{inf}} = 10^{9}\, \mathrm{GeV}$),
the constraint on the curvaton-like behavior (upper left boundaries),
and the constraint on the baryon
isocurvature arising from deviations from the slow-varying trajectory
(lower boundaries for $H_{\mathrm{inf}} = 10^{10}, \, 10^{11}\, \mathrm{GeV}$).
Comparing with Figure~\ref{fig:frac1} of the $n = 1$ case, one sees that
the parameter window is now narrower, which can also be seen from
comparing the powers of $\lambda$ in the conditions 
(\ref{cond-ii-n2}), (\ref{r_for_n2}) with the corresponding 
(\ref{cond-ii}), (\ref{2.30}).
(Note also the dependence of $T_{\mathrm{dec}}$ on~$\lambda$ after
fixing the baryon-to-entropy ratio (\ref{3.4}), (\ref{2.21}).)
When the ratio $\lambda/f$ is smaller than~$10^{-1}$,
the window for $n=2$ becomes much narrower than for $n = 1$. 

We have also estimated the cutoff of the theory 
by power counting.
For the parameters in the figure,
both the derivative coupling and the $Z_2$-breaking terms 
give a cutoff in the large field regime during baryogenesis as $\Lambda
\sim \sigma_\star$.
The condition $T_{\mathrm{dec}} < \Lambda$ is satisfied on the left
side of the black dashed line, hence one sees that $T_\mathrm{dec}$ is
below the cutoff in the entire allowed window.

\section{Conclusions}
\label{sec:conc}

In this work, we proposed a new class of spontaneous baryogenesis models
that does not produce baryon isocurvature perturbations during the
generation of the baryon asymmetry.
The basic idea is that when the baryon-generating scalar possesses a
potential and a coupling satisfying the condition~(\ref{general}),
the induced baryon chemical potential is independent of the
scalar field and thus becomes spatially homogeneous.\footnote{One may
wonder whether a similar 
technique could also be used to suppress isocurvature perturbations of
scalar field dark matter, such as axions.
In order to achieve this, the field fluctuation should be blocked from
producing fluctuations in the dark matter density~$V(\sigma)$, which seems
difficult unless there is overtaking such that fields starting higher
up in the potential reaches the potential minimum faster.} We
demonstrated this mechanism in models that involve non-canonical 
scalar fields with actions of the form~(\ref{Ln})
which, after canonical normalization, reduces 
to a theory satisfying~(\ref{general}).
The cosmological history in these non-canonical models is similar to
that of vanilla spontaneous baryogenesis, except for that the scalar
field undergoes anharmonic oscillations along a non-quadratic
potential and hence its density redshifts in a specific way. We analyzed
in detail the baryon generation and the cosmology for the
models of $n = 1$ and $2$, and investigated the parameter space that
gives rise to successful baryogenesis.
We found that our models allow scalar masses and couplings, as well as
decoupling and inflation scales, in regions that can be quite different from those in
vanilla scenarios (see e.g. \cite{DeSimone:2016ofp} for comparison).
In particular, the suppression of the baryon isocurvature allows
spontaneous baryogenesis with inflation scales up to
$H_{\mathrm{inf}} \sim 10^{14}\, \mathrm{GeV}$.
This implies that, even if inflationary gravitational waves are detected
in the near future, 
spontaneous baryogenesis would not be ruled out.

Clearly an important direction for further study is to realize the
setup for isocurvature suppression within fundamental physics constructions. 
The Nambu--Goto action of branes could serve as a potential candidate
for realizing the non-canonical kinetic term,
as in the  D-brane monodromy inflation models
of~\cite{Silverstein:2008sg,McAllister:2008hb}.
We should also note that the runnings of the couplings may affect
the condition~(\ref{general}).
In this paper we have discussed the validity of the effective field
theory using power 
counting arguments, however it is also important to analyze in more detail
the stability of the theory against radiative corrections.
We leave these investigations for future work.

Although we have focused on non-canonical scalar fields, the
mechanism of suppressing the baryon isocurvature is much more general.
It would also be interesting to explore other ways to realize the
scalar-independent chemical potential of~(\ref{general}).
We stress that in these models without baryon isocurvature,
the resulting baryon asymmetry is independent of the scalar field value. 
This gives predictive power to the model,
especially in cases where the baryon-generating scalar arises as a
Nambu--Goldstone boson of a spontaneously broken symmetry and thus has
random initial conditions.

\section*{Acknowledgments}

It is a pleasure to thank Andrew Cohen and Fuminobu Takahashi for very
helpful discussions.
T.K. acknowledges support from the INFN INDARK PD51 grant.





\end{document}